\documentclass[12pt]{article}

\newcommand{\real}{\mathrm{Re~}}
\newcommand{\imag}{\mathrm{Im~}}
\newcommand{\Tr}{\mathrm{Tr~}}

\newcommand{\id}{\mathbbm{1}}

\usepackage{bm,bbm,graphicx,amssymb}
\usepackage{hyperref}

\makeatletter

\@addtoreset{equation}{section}

\@addtoreset{figure}{section}
\makeatother

\title{\bf Vortex Description of Quantum Hall Ferromagnets}

\author{Taro Kimura\thanks{E-mail: kimura@dice.c.u-tokyo.ac.jp}\\
\vspace{1em}\\
\small{\it Department of Basic Science, University of Tokyo,}\\
\small{\it Meguro-ku, Komaba, Tokyo 153-8902, Japan}
}

\date{}

\begin{document}
\maketitle

\begin{abstract}
We study particle states of quantum Hall ferromagnet from the
 viewpoint of the incompressible fluid description.
It is shown that phase space of Chern-Simons matrix theory which is an
 effective theory for the incompressible fluid is equivalent to moduli
 space of vortex theory.
According to this correspondence, 
elementary excitations in vortex theory are identified as particle
 states in quantum Hall ferromagnet, and thus we propose that a pure
 electron state is absent from the strong coupling region but only a
 composite particle state is present.
\end{abstract}

\newpage


\section{Introduction}

The quantum Hall effect is one of the most remarkable phenomena in
condensed matter physics, and
gives a rich mathematical structure\cite{prange1990qhe}. 
It is well known that electrons in a low energy region behave as
incompressible fluid. 
An important property of the incompressible fluid is that it possesses
no dynamical degree of freedom and the residual degree of freedom comes
from geometry of the fluid, which is related to area preserving diffeomorphism.
Thus Chern-Simons theory which is also non-dynamical theory captures the
feature of the incompressible fluid. 
Indeed one can derive Chern-Simons action by integrating out fermion modes\cite{PhysRevB.44.5246}.
This situation is similar to topological string theory omitting a
string fluctuation, but mainly treating its topological structure. 

Then we remark that the relation between incompressibility and
noncommutativity. 
The most primitive example of the noncommutativity is the canonical
commutation relation $[x,p]= i \hbar$. 
According to this noncommutativity, quantum mechanical phase space
becomes fuzzy, and a quantum state covers an area $\sim \hbar$, which is
called a quantum droplet, and is preserved while its shape is
transformed. 
On the other hand, a classical mechanical state is indicated by a point on
classical phase space. 
In the case of the magnetic system, a momentum $\vec{p} = m \dot{\vec x}
- \vec{A}$ includes a coordinate component via the vector potential
$\vec A$. 
Therefore the coordinate space, which is the phase space itself, becomes
noncommutative and a state of particles is interpreted as the
incompressible fluid. 
In general, the noncommutativity of the space-time is induced by the
effect of background fields\cite{seiberg1999sta}. 

To manifest the noncommutativity of the quantum Hall state, the
noncommutative analogue of Chern-Simons theory was proposed as the
effective theory of the incompressible fluid\cite{susskind2001qhf}. 
Since the canonical commutation relation can be realized by infinite
dimensional matrices, the corresponding system is infinitely extended
without the boundary. 
Then the regularized finite model, the Chern-Simons matrix model was
also presented\cite{polychronakos2001qhs,polychronakos2001uqhs}, 
in which the commutation relation is modified by the boundary effect.
On the other hand, the same relation was discovered in the context of vortex
theory, in particular the usual quantum Hall state corresponds to the
Abelian vortex state\cite{hanany2003via}. 
As a result, the phase space of the quantum Hall state turns out to be
identified with the moduli space of the vortex theory\cite{tong2004qhf}. 

In this paper, we investigate non-Abelian generalizations of the correspondence
between the quantum Hall state and the vortex theory.
When two dimensional layers are stacked, we can consider the internal
degree of freedom labeling the layers.
It is well known as the pseudo spin. 
The quantum Hall state with the spin or pseudo spin degree of freedom is
called a quantum Hall ferromagnet.  
The enhanced $SU(2)$ symmetry is decomposed to the electron part $U(1)$
and the spin part $SU(2)/U(1) = \mathbb{C}{\bf P}^1$, and this is
interpreted as the spin-charge separation. 
Hence a spin wave as the Nambu-Goldstone mode is induced by the
spontaneous symmetry breaking, and we can observe a skyrmion which is a
topological excitation characterized by a non-trivial homotopy class
$\pi_2[SU(2)/U(1)]=\mathbb{Z}$. 

The $\mathbb{C}{\bf P}^1$ space is also obtained as the internal space
of the non-Abelian vortex. 
Thus, identifying these internal spaces, we can investigate the quantum
Hall ferromagnet by the vortex theory 
and apply the $\mathbb{C}{\bf P}^1$ valued field theory, especially
${\cal N}=(2,2)$ supersymmetric $\mathbb{C}{\bf P}^{1}$
model\cite{dorey1998bst,shifman2006cca} to vortex world-sheet
theory\cite{Hanany:2004ea}. 
This model has been studied in the context of the mirror
symmetry\cite{Hanany:1997vm} and also applied to
superconductivity\cite{hikami2007ftc}. 

The structure of this paper is the following.
In section \ref{sec:QHE}, we review the relationship between the
noncommutativity and the incompressibility of the quantum Hall state and
the effective theory of the incompressible fluid for the finite system. 
In section \ref{sec:vortex}, it will be shown that the regularized
commutation relation of the incompressible fluid is also obtained from
vortex theory and the moduli space of vortices is identified with
the phase space of the incompressible fluid.
Due to the internal symmetry of the vortex, we introduce ${\cal
N}=(2,2)$ supersymmetric $\mathbb{C}{\bf P}^{1}$ theory describing
internal particle state of the quantum Hall ferromagnet.
In section \ref{sec:CP}, particle states of the quantum Hall ferromagnet
are investigated by the vortex theory.
In the $\mathbb{C}{\bf P}^1$ theory, there exist two phases, which are
strong and weak coupling phase, and they are
separated by the curve of marginal stability.
The strong coupling phase is considered as the quantum Hall ferromagnet
state, but the weak coupling phase is not.
Subsequently, we propose that only a composite particle state appears
as a charged particle but a pure electron state is absence from the
quantum Hall ferromagnet.


\section{Quantum Hall effect and noncommutativity}\label{sec:QHE}

In this section we review the derivation of the noncommutative Chern-Simons theory as the effective theory of the incompressible Quantum Hall fluid, based on \cite{susskind2001qhf,polychronakos2001qhs,polychronakos2001uqhs}.

In a two dimensional system with perpendicular magnetic field,
cyclotron motion of an electron is quantized as the harmonic oscillator,
and discretized energy levels are called Landau levels.
The density of states in the lowest Landau level (LLL) is uniform and
in proportion to the strength of the magnetic field,
\begin{equation}
 \rho_0 = \frac{1}{2\pi l_0^2}
\end{equation}
where $l_{0}=1/\sqrt{B}$ is the magnetic length characterizing the scale of
the wave function, and thus almost all electrons fall into the LLL in
strong magnetic limit. 
Since the density is spatially constant, occupied area is exactly determined
by fixing the number of particles.
While the area is preserved, positions of particles can be changed by
gauge transformation. 
Therefore, the electron state in the strong magnetic field behaves as
incompressible fluid.
Although any dynamical degrees of freedom do not exist because
we neglect excitations to higher Landau levels, we should
consider residual degrees of freedom for the fluid, geometrical configurations of
particles, related to area preserving transformation.

The effective theory of the LLL state is often derived by integrating
out fermion modes\cite{PhysRevB.44.5246}.
In this paper, we show another way to obtain the effective theory.
We firstly introduce integration
constants of the cyclotron motion describing the residual
degrees of freedom called a guiding center: 
\begin{equation}
 X = x + l_{0}^2 \Pi_y, \qquad
 Y = y - l_{0}^2 \Pi_x
\end{equation}
where $\vec{\Pi}=\vec{p}+\vec{A}$ is the
magnetic momentum.
These operators satisfy the following commutation relations
\begin{equation}
 \left[ X, Y \right] = i l_{0}^2, \qquad
 \left[ \Pi_x, \Pi_y \right] = - \frac{i}{l_{0}^2}. \label{NC01}
\end{equation}
This spatial noncommutativity is considered as an example of correspondences
between commutative theory with background 
field and noncommutative theory which is well known as
Seiberg-Witten map\cite{seiberg1999sta}.

When the magnetic field becomes so strong, contributions of the magnetic
momentum to the guiding center and the canonical momentum can be
neglected as $\vec{X}\simeq \vec x$ and $\vec{p} \simeq -\vec{A}$.
Thus the Lagrangian can be written in terms of the guiding center coordinates
\begin{equation}
 {\cal L} = \vec{p}\cdot\vec{\dot x} - {\cal H} =
  \frac{B}{2}\left(X \dot{Y} - \dot{X} Y\right), \qquad {\cal H} =
  \frac{1}{2m}\left|\vec{\Pi}\right|^2. \label{eff_lagrangian}
\end{equation}
This Lagrangian induces only the Lorentz force and mechanical work can
be zero.
Since our theory is not dynamical, we can generalize
(\ref{eff_lagrangian}) to the $n$-body state action
\begin{equation}
 {\cal S} = \frac{B}{2} \int dt~ \sum_{\alpha=1}^n \epsilon_{ab}
  X^a_\alpha \dot{X}^b_\alpha \label{NCCS01}
\end{equation}
where $X^1=X$, $X^2=Y$ and the subscript $\alpha = 1, \cdots,
n$, is an index for particles. 
In the large $n$ limit, a fluid dynamical description becomes available
\begin{equation}
 \sum_{\alpha=1}^n \to \int d^2 x~ \rho(\vec{x}), \qquad
 \vec{X}_\alpha (t) \to \vec{X}(\vec{x},t), \qquad
 \vec{X}(\vec{x},0) = \vec{x}.
\end{equation}
The initial state is a reference configuration of the fluid. 
We will consider fluctuation modes from the reference state as the
residual degree of freedom.

The constraint for the incompressibility is the constant density
condition, $\rho(\vec{x})=\rho_e$.
Since the density of particles is the Jacobian of the fluid dynamical
field, the constraint can be written with Poisson bracket form
\begin{equation}
 \rho_e = \rho(\vec{x}) = \rho_e \left| \frac{\partial \vec{X}}{\partial
				  \vec{x}} \right| = \frac{1}{2}\rho_e
 \epsilon_{ab} \left\{X^a, X^b\right\}.
\end{equation}
Adding this Jacobian preservation constraint to (\ref{NCCS01}) with
temporal gauge field $A_0$ as the Lagrange multiplier, the
action is modified as 
\begin{equation}
 {\cal S} = \frac{B}{2} \rho_e \int dt\ d^2 x \left[ \epsilon_{ab} X^a \left(
     \dot{X}^b - \theta \left\{X^b, A_0\right\}\right)+ 2\theta
 A_0\right] \label{NCCS02}
\end{equation}
where $\theta = 1/(2\pi \rho_e)$ will become the noncommutative parameter.
Then, satisfying the constraint, we can decompose $X^a$ as 
\begin{equation}
 X^a = x^a + \theta^{ab} A_b, \qquad \theta^{ab} = \theta \epsilon^{ab}.
\end{equation}
Here we can regard gauge fields as the fluctuation mode from the
reference state, and the gauge transformation corresponds to area
preserving transformation of the fluid. 
Writing the action (\ref{NCCS02}) in terms of the gauge fields, we
obtain
\begin{equation}
 {\cal S} = \frac{1}{4\pi\nu} \int dt\ d^2 x\ \epsilon^{\mu\nu\lambda}
  \left( \partial_\mu A_\nu A_\lambda + \frac{\theta}{3} \left\{A_\mu,
							  A_\nu\right\}A_\lambda\right).
\end{equation}
The constant $1/\nu=1/(B\theta)$ is an integer, which is the level of
the Chern-Simons theory, 
and $\nu=\rho_e/\rho_0$ is a filling fraction for the LLL states.
Furthermore, this action can be regarded as a leading contribution of
noncommutative Chern-Simons action\cite{susskind2001qhf}
\begin{equation}
 {\cal S}_{\mathrm{NCCS}} = \frac{1}{4\pi\nu} \int dt\ d^2 x\
  \epsilon^{\mu\nu\lambda} \left( \partial_\mu A_\nu \star A_\lambda -
			    \frac{2}{3}i A_\mu \star A_\nu \star A_\lambda \right)
\end{equation}
where $\star$-product is the Moyal product defined as
\begin{equation}
 f(x) \star g(x) = f(x) \exp \left( \frac{i}{2}
			  \stackrel{\leftarrow}{\partial_\mu}
			  \theta^{\mu\nu}
			  \stackrel{\rightarrow}{\partial_\nu} \right) g(x).
\end{equation}
Because this product is noncommutative, the commutation relation is
naively modified
\begin{equation}
 \left[ x_1, x_2 \right]_\star = x_1 \star x_2 - x_2 \star x_1 = i \theta. 
\end{equation}
This noncommutativity is analogous to (\ref{NC01}).
This means that the noncommutative relation for one particle state is
generalized to multi-particle fluid state.


The noncommutative relation can be also represented by
regarding $X^a$ as an infinite dimensional matrix acting on Hilbert space.
The corresponding matrix model becomes Chern-Simons matrix model
\begin{equation}
 {\cal S}_{\mathrm{MCS}} = \frac{B}{2} \int dt\ \Tr \left[ \epsilon_{ab}
						    X^a \left( \dot{X}^b
							- i \left[A_0, X^b
							    \right]\right)
						    + 2 \theta A_0\right].
\end{equation}
The spatial integration is replaced with taking the matrix trace.
Then we immediately obtain the equation of motion for the non-dynamical
variable $A_0$ as
\begin{equation}
 \left[X^1, X^2\right] = i\theta \id_\infty. \label{NC02}
\end{equation}
Here the right hand side of (\ref{NC02}) is in proportional to the
infinite dimensional identity matrix.
That means this action is well defined only when the number of particles is
infinite.
However natural quantum Hall states are realized with the finite
system where the boundary state plays an essential role on the transport
phenomena. 
To regularize the infiniteness of the Hilbert space, one should introduce a
boundary field which seems to correspond to the edge state.
Thus we obtain a regularized finite matrix model proposed in \cite{polychronakos2001qhs,polychronakos2001uqhs},
\begin{equation}
  {\cal S}_{\mathrm{MCS}} = \frac{B}{2} \int dt~ \Tr \left[
							  \epsilon_{ab}
							  X^a \left(
							       \dot{X}^b
							     - i
							     \left[A_0,
							      X^b\right]\right)
							 + 2\theta A_0 -
							\omega
							\left(X^a\right)^2\right]
 + \int dt~ \Psi^\dagger \left(i \dot{\Psi} - A_0 \Psi\right).
\end{equation}
The quadratic term $\omega \left(X^a\right)^2$ is the confinement
potential and $\Psi$ is $n$ component bosonic field absorbing boundary anomaly.
Thus the equation of motion for the Lagrange multiplier $A_0$ is obtained as
\begin{equation}
 \left[ X^1, X^2 \right] = i \theta \id_n - \frac{i}{B} \Psi
  \Psi^\dagger \label{NC03}
\end{equation}
with the normalization condition, $\Psi^\dagger \Psi = nB\theta = n/\nu$.
In this case, the modified commutation relation (\ref{NC03}) is
realized with $n\times n$ matrices $X^a$.
Introducing a complex matrix $X=\left(X^1 + i X^2\right)/\sqrt{2}$ and
$X^\dagger = \left(X^1 - i X^2\right)/\sqrt{2}$, the noncommutative relation
(\ref{NC03}) is rewritten as
\begin{equation}
 \frac{1}{B} \Psi \Psi^\dagger + \left[ X, X^\dagger \right] -
  \theta \id_n = 0. \label{NC04}
\end{equation}
The number of parameters for the physical phase space satisfying this
constraint, the dimension of the phase space, is $2n^2 + 2n - 2n^2 = 2n$.
Thus these parameters can be regarded as two dimensional coordinates
of particles. 
This relation is of quantum Hall state without internal degrees of
freedom.
In the following section we will see this relation also
appears in the vortex theory, and thus its non-Abelian generalization is
considered.

\section{Vortex theory and quantum Hall ferromagnets}\label{sec:vortex}

Topological excitations, e.g. vortices, instantons, play an important
role on non-perturbative aspects of quantum field theory.
Although solutions of $k$-instanton with arbitrary $k$ was
constructed in \cite{atiyah1978cip}, an explicit vortex solution is not
yet found. 
However, the structure of the vortex moduli space was recently
conjectured by the stringy method\cite{hanany2003via}.
In this section, we start with a review of the vortex moduli space based on \cite{hanany2003via}. We then discuss a relationship to the incompressible fluid, and show that quantum Hall state is
considered as vortex fluid state.

\subsection{Vortex moduli space}

We want to investigate vortices in $2+1$ dimensional ${\cal N}=4$
supersymmetric gauge theory.
The $U(N)_G$ vector multiplet consists of a gauge field $A_\mu$, a triplet
of adjoint scalar fields $\phi^r$, and fermionic partners.
The $N$ fundamental hypermultiplets are complex scalars $q$, $\tilde q$
and fermions. 
Furthermore, considering $SU(N)_F$ flavor symmetry, fundamental
fields $q$ and $\tilde q$ obey $(N,\bar N)$ and $(\bar N,N)$
representation respectively.
Thus we write the bosonic part of the Lagrangian of this theory
\begin{eqnarray}
 {\cal L} & = & - \Tr \Big[ \frac{1}{2e^2} F_{\mu\nu} F^{\mu\nu} +
  \frac{1}{2e^2} D_\mu\phi^r D^\mu\phi^r + D_\mu q^\dagger D^\mu q +
  D_\mu \tilde q D^\mu \tilde q^\dagger + e^2 \left|q\tilde q\right|
  \nonumber \\ 
 & & \quad + \frac{1}{2e^2} \left|\left[\phi^r,\phi^s\right]\right|^2 + \left(\tilde q^\dagger \tilde q - qq^\dagger\right)\phi^r
  \phi^r + \frac{e^2}{2}\left(qq^\dagger - \tilde q^\dagger \tilde q -
			 \zeta \id_N\right)^2 \Big] \label{2+1dim_lagrangian}
\end{eqnarray}
where $\zeta$ is the Fayet-Iliopoulos (FI) parameter which ensures the
symmetry broken vacuum.
The ground state of this model is gapped, and then vortices appear with
the broken symmetry
\begin{equation}
 U(N)_G \times SU(N)_F \quad \longrightarrow \quad SU(N)_{\mathrm{diag}}.
\end{equation}

Then we construct this model by D-brane configuration with $N$
D3-branes and $k$ D1-branes which are regarded as the space-time and vortices
respectively. 
In the decoupling limit of the string fluctuation, dynamics of
D1-branes can be described by ${\cal N}=(2,2)$ supersymmetric
quantum mechanics\footnote{When we consider $4$ dimensional ${\cal
N}=2$ supersymmetric theory, vortex theory becomes $1+1$
dimensional ${\cal N}=(2,2)$ supersymmetric theory.}.
In this model, $U(k)$ vector multiplet consists of a gauge field and
adjoint scalars $\phi^r$ corresponding to vortex fluctuations of
perpendicular directions.
Thus two dimensional positions of vortices are described as a complex
scalar $Z$ of the adjoint chiral multiplet.
The fundamental chiral multiplets, complex scalars $\psi$, come from
excitations of D1-D3 strings. 
Then the bosonic Lagrangian on D1-branes becomes
\begin{eqnarray}
 {\cal L}_{\mathrm{vortex}} & = & \Tr \Big[ \frac{1}{2g^2}D_t \phi^r D_t
  \phi^r + D_t Z^\dagger D_t Z + D_t \psi^i D_t \psi^\dagger_i -
  \frac{1}{2g^2}\left[ \phi^r, \phi^s\right]^2  \nonumber \\
 && \qquad - \left|\left[Z,\phi^r\right]\right|^2 - \psi^i \psi^\dagger_i \phi^r \phi^r - \frac{g^2}{2} \left( \psi^i
							     \psi^\dagger_i
							     -
							     \left[Z,Z^\dagger\right]
							     - r
							     \id_{k}\right)^2 \Big].
\end{eqnarray}
The FI parameter of this model is identified with the original gauge
coupling as $r = 2\pi/e^2$.
For finite $r\not=0$, we should consider Higgs branch in the
decoupling limit $g^2\to\infty$, and thus $k\times k$
$D$-term condition reads 
\begin{equation}
 \psi^i \psi^\dagger_i - \left[Z, Z^\dagger\right] - r \id_k = 0 \label{NC05}
\end{equation}
where $i$ is the index of the gauge group $U(N)_G$ running as $i=1, \cdots, N$.
The number of parameters of the Higgs branch is $2kN+2k^2-2k^2=2kN$ since
$Z$ and $\psi$ are $k\times k$ and $k\times N$ matrices.
These matrix valued fields parametrize positions of vortices.
Such kinds of parameters for the solution space are called moduli,
 and the corresponding parameter space is called moduli space.
Therefore we can regard this Higgs branch as the moduli space ${\cal
M}_{k,N}$ for $k$-vortex solution with $U(N)_G$ gauge symmetry.

Adjusting some normalizations, the noncommutative relation for the
incompressible fluid (\ref{NC04}) is equivalent to the Abelian ($N=1$) case
of the vortex relation (\ref{NC05}) when we identify the number of
particles $n$ with the vortex number $k$ and the noncommutative parameter $\theta$ with the FI parameter $r$.
This means that we can regard particles of the incompressible fluid as
Abelian vortices.
In this aspect, the geometry of the Abelian vortex moduli space was
discussed in \cite{tong2004qhf}.
In fact, since the vortex width $l_v$ is evaluated as $l_v \sim \sqrt{r}$, the
particle density becomes $\rho_e \sim 1/(2\pi l_v^2) \sim 1/(2\pi r)$.
This estimation is consistent with our identification $r \sim \theta$.

Due to this relation, we want to consider incompressible fluid
consisting of non-Abelian vortices.
Indeed quantum Hall state with internal symmetry is known as a
quantum Hall ferromagnet and its internal degree corresponds to not only
spin of a particle but an index of multilayer systems, which
is called a pseudo spin.

To discuss the relationship between quantum Hall
state and vortex theory, we investigate the moduli space of vortices.
From the vortex relation (\ref{NC05}), the moduli space of 1-vortex
state is determined,
\begin{equation}
 {\cal M}_{1,N} \cong \mathbb{C} \times \mathbb{C}{\bf P}^{N-1}.
\end{equation}
This means that the 1-vortex moduli is decomposed to a position of vortex
center $\mathbb{C}$ and internal $\mathbb{C}{\bf P}^{N-1}$ space.
On the other hand, the moduli space for $k$-Abelian vortex is obtained
by \cite{Taubes:1979tm} as
\begin{equation}
 {\cal M}_{k,1} \cong \mathbb{C}^k / \mathfrak{S}_k
\end{equation}
where $\mathfrak{S}_k$ is symmetric group.
Then the higher $k$ moduli space is also represented as symmetric
product\cite{Eto:2005yh} 
\begin{equation}
{\cal M}_{k,N} \cong \left( \mathbb{C} \times \mathbb{C}{\bf
	      P}^{N-1}\right)^k / \mathfrak{S}_k,
\end{equation}
and it is suggested that the orbifold
singularity of vortex collision is smoothed
out\cite{Eto:2005yh,Auzzi:2005gr,eto:065021}. 


We now consider non-Abelian vortex fluid state as quantum Hall state with internal symmetry.
In fact, for $N$-layered quantum Hall state, each particle has $SU(N)$
symmetry, but its $U(1)$ part is decoupled as electromagnetic part.
Thus the residual $\mathbb{C}{\bf P}^{N-1}$ part is interpreted as internal
symmetry of a particle.
This phenomenon is called spin-charge separation.
Although explicit derivation of non-Abelian generalization of (\ref{NC04}) has not been found, 
relying on the coincidence of the internal symmetry, we identify quantum Hall ferromagnets with non-Abelian vortex fluid.

Therefore, according to the supersymmetry of the original field theory,
we choose $1+1$ dimensional ${\cal N}=(2,2)$ supersymmetric
$\mathbb{C}{\bf P}^{N-1}$ model for the vortex world sheet
theory\cite{Hanany:2004ea}. 
Although the physical meaning of superpartners in quantum Hall ferromagnets is not clear, since (\ref{NC05}) is derived from $D$-term condition of the supersymmetric theory, we apply supersymmetric effective theory to the vortex fluids.
However, some properties of solitons in supersymmetric theory are actually observed in the incompressible fluid.
We then explain them as follows.

The well known vortical model describing the
superconductor, which is called the Ginzburg-Landau model, has two
independent coupling constants, the electromagnetic constant $e$ and the
condensate coupling $\lambda$,
\begin{equation}
 {\cal L}_{\mathrm{GL}} = - \frac{1}{2e^2} F_{\mu\nu}F^{\mu\nu} + D_\mu
  \phi D^\mu \phi^\dagger - \frac{\lambda}{2}\left(\phi \phi^\dagger - \zeta\right)^2.
\end{equation}
Each parameter corresponds to characteristic lengths of the
superconductor, coherence length and penetration length. 
Thus the type of the superconductor is determined by the ratio of these
lengths called Ginzburg-Landau parameter, $\kappa = \sqrt{\lambda/(2e^2)}$.
For $\kappa \ll 1$, the superconductor is of type I in which the interaction
between vortices are attractive.
For $\kappa \gg 1$, it becomes the type II superconductor where vortices
are repulsive. 
In the case of (\ref{2+1dim_lagrangian}), since our superconductor $\kappa=1/\sqrt{2}$ is
the intermediate state of type I and II, the vortices become
interactionless. 
This is one of the features of solitons in supersymmetric gauge theory.
On the other hand, we now remark interactions between these solitons arise when they are moving.
As discussed in section \ref{sec:QHE}, Chern-Simons theory
captures the geometric property of the incompressible fluid when its
dynamics is neglected.
This means we consider interactionless particles, namely, the static sector of the LLL.

\subsection{Vortex world-sheet theory}\label{world_sheet}

Then we consider the field theory describing the dynamics of
the non-Abelian vortex.
To discuss the quantum Hall ferromagnets, we now give a brief review of
the supersymmetric $\mathbb{C}{\bf P}^{N-1}$ model based on
\cite{dorey1998bst,shifman2006cca,Hanany:1997vm}. 

To consider the supersymmetric generalization of the bosonic model, it is
convenience to introduce the superfield formulation\footnote{Not to be confused with the fermionic parameter $\theta$ and the noncommutative
parameter $\theta=1/(2\pi \rho_e)$.}.
Chiral and anti-chiral superfield are defined as $\Phi^j (x^\mu + i \bar \theta \gamma^\mu \theta)$ and $\Phi^{^\dagger \bar j} (x^\mu - i \bar \theta \gamma^\mu \theta)$.
Thus $D$-term Lagrangian of the supersymmetric $\mathbb{C}{\bf P}^{N-1}$
model is written as
\begin{eqnarray}
 {\cal L} & = & \int d^4 \theta\ K(\Phi, \Phi^\dagger) \nonumber \\
 & = & G_{i\bar j} \left[ \partial^\mu \phi^{\dagger \bar j}
		    \partial_\mu \phi^i + i \bar \psi^{\bar
		    j}\gamma^\mu D_\mu \psi^i - \frac{1}{2}R_{i\bar j k
		    \bar l}\left(\bar\psi^{\bar
			    j}\psi^i\right)\left(\bar \psi^{\bar l}\psi^k\right)\right]
\end{eqnarray}
with K\"ahler metric $G_{i\bar j}=\partial_i \partial_{\bar{j}}K$,
Riemann tensor $R_{i\bar j k \bar l}$ and covariant derivative $D_\mu$.
In the case of the $\mathbb{C}{\bf P}^{N-1}$ model, the K\"ahler potential is defined as
\begin{equation}
 K(\Phi, \Phi^\dagger) = \frac{2}{g^2} \log \left( 1+\Phi^\dagger
					       \Phi\right). \label{Kahler_pot01}
\end{equation}
Although the fields in this model are different from those in the
previous section, the coupling constant $g^2$ is same as that of the
corresponding gauge theory.
Actually, at the low energy region, it goes to infinity $g^2 \to
\infty$, and is consistent with the decoupling limit of string fluctuation.

For the two dimensional supersymmetric theory, spinor irreducible
representation is Majorana-Weyl type, and we can include a twisted chiral field
in addition to the chiral field.
Thus the K\"ahler potential with twisted masses $m_a$ which are the
classical vacuum expectation values of the twisted chiral
superfield, $\Sigma = \sigma + \sqrt{2}\theta\tilde \chi + \theta^2 S$, is obtained by modifying the usual K\"ahler potential (\ref{Kahler_pot01}) for $\mathbb{C}{\bf P}^{N-1}$ manifold,
\begin{equation}
 K(\Phi, \Phi^\dagger, V) = \frac{2}{g^2}
 \log \left(1+\Phi^\dagger e^{V_aT^a}   \Psi\right)
\end{equation}
where $\left( T^a \right)^i_j=\delta^i_a \delta^a_j \ (a=1, \cdots, N-1)$ are the generators with a diagonal form, and corresponding external $U(1)$ components are written as a complex form,
\begin{equation}
 V_a = -m_a \bar \theta (1+\sigma_3) \theta - \bar m_a \bar \theta (
  1-\sigma_3) \theta,
\end{equation}
\begin{equation}
 m_a = A_y^a + i A_x^a , \quad \bar m_a = m_a^* = A_y^a - i A_x^a.
\end{equation}
Here $\sigma_3$ is Pauli matrix and we can set $\sum_{a=1}^N m_a=0$ by
shifting the twisted chiral field.

Then let us discuss the vacuum structure of this model.
The low energy effective action is obtained by integrating out the
chiral superfield and written in terms of the twisted chiral field.
To consider the additional contribution of $F$-term corresponding to
FI-term and $\vartheta$-term, we introduce the twisted superpotential at
classical level,
\begin{equation}
 {\cal W} = \frac{i}{2} \tau \Sigma
\end{equation}
where $\tau$ is a complex coupling constant obtained by introducing a theta angle $\vartheta$,
\begin{equation}
 \tau = \frac{2i}{g^2} + \frac{\vartheta}{2\pi}.
\end{equation}
Although there exists only one classical vacuum at $\Sigma=0$, we will show the quantum vacuum possesses more rich structure.
The dynamically generated mass is exactly evaluated by the renormalization group equation at one loop order with a reference point $\mu$,
\begin{equation}
 \Lambda = \mu e^{-\frac{4\pi}{Ng^2}}. \label{dynamical_mass}
\end{equation}
The twisted superpotential with the twisted masses is also corrected by
the renormalization effect.
Thus the effective potential is given by
\begin{equation}
 \tilde{\cal W} = \frac{i}{2} \left[ \tau \Sigma - \frac{1}{2\pi i}\sum_{a=1}^N \left(\Sigma - m_a \right) \log \left( \frac{2}{\mu} \left( \Sigma - m_a \right) \right) \right].
\end{equation}
In this case, vacua of this potential can be determined by differentiating with the twisted chiral field
\begin{equation}
 \frac{\partial \tilde{\cal W}}{\partial \Sigma} = 0 
 \quad \longrightarrow \quad
 \prod_{a=1}^N \left( \sigma - m_a \right) - \tilde \Lambda^N = 0
\end{equation}
where $\tilde \Lambda = (\mu/2) \exp \left( 2\pi i \tau / N -1 \right)$ is a complexified dynamical mass.
This condition ensures that there exist $N$ vacua in the quantum level,
and then we can consider a topological kink solution.

In the case of $N=2$ theory, which corresponds to
$\mathbb{C}{\bf P}^1$ model, the renormalization point $\mu$ can be
replaced with the twisted mass $m$, with $m_2=-m_1=m/2$.
The mass of the kink solution becomes
\begin{equation}
 m_D = \tilde{\cal W}(\sigma_+) - \tilde{\cal W}(\sigma_-) = -
  \frac{i}{2\pi}\left[m\log\left(\frac{m-\sqrt{m^2+4\tilde\Lambda^2}}{m+\sqrt{m^2+4\tilde\Lambda^2}}\right)+2\sqrt{m^2+4\tilde\Lambda^2}\right] \label{center01}
\end{equation}
where $\sigma_{\pm}=\pm \sqrt{m^2/4+\tilde\Lambda^2}$ are solutions of 
\begin{equation}
 \sigma^2 - \frac{m^2}{4} - \tilde \Lambda^2 = 0.
\end{equation}
From the mass of the topological excitation $m_D$ which is exactly evaluated in (\ref{center01}) and the elementary mass $m$, we obtain the central charge of $1+1$ dimensional superalgebra $Z=n_e m + n_D m_D$ which characterizes the BPS mass $M=|Z|$.

To discuss the BPS states in the strong coupling region
$\left|m^2/4\tilde\Lambda^2\right|\ll 1$ and the weak coupling region
$\left|m^2/4\tilde\Lambda^2\right|\gg 1$, we expand the topological mass
in terms of the mass parameter,
\begin{equation}
 m_D = \frac{im}{\pi} \left[ i\pi + \log\left(\frac{m}{\tilde\Lambda}\right) + \sum_{k=1}^\infty c_k \left(\frac{\tilde\Lambda}{m}\right)^{2k} \right]
\end{equation}
where $c_k = (-1)^k (2k-2)! / (k!)^2$.
The first term is the tree level contribution, and the second is the one loop correction.
The infinite series of the last term comes from the instanton effect.

In the weak coupling limit $\left|m^2/4\tilde\Lambda^2\right|\gg 1$, since the ratio of two masses increases logarithmically, $m_D/m \sim \log m$, the topological excitation is restricted, and thus surviving BPS states are $n_e = \pm 1$, $n_D=0$ and $n_D=\pm 1$ with arbitrary $n_e$.
On the other hand, the situation at strong coupling $\left|m^2/4\tilde\Lambda^2\right|\ll 1$ is similar to the usual $\mathbb{C}{\bf P}^{1}$ model in the absence of the twisted mass.
When we shift the theta angle $\vartheta \to \vartheta + 2\pi$, a sign of the mass is inverted $m \to -m$.
This means that a relevant parameter is the squared mass $m^2$ and we
have a cut singularity along negative part of the real axis of the
complex $m^2$ plane.
The range of the cut is $[-1,0]$.
Therefore the monodromy around infinity on the $m^2$ plane is obtained as $(m, m_D) \to (-m, -m_D - m)$, which is equivalently $(n_e, n_D) \to (-n_e + n_D, -n_D)$.
As a result, the BPS states in the strong coupling region are only $(n_e, n_D)=(0,1)$, $(1, -1)$ and their anti-excitations.

Since the structures of the BPS states in the strong and weak coupling regions are quite different, they must be separated by a curve of marginal stability (CMS) on the complex $m^2$ plane.
The CMS is obtained as a coincidence condition of phases of the elementary mass and the topological mass, simply written as
\begin{equation}
 \imag \left( \frac{m_D}{m} \right) = \real \left[\log\left(\frac{1-\sqrt{1+4\tilde\Lambda^2/m^2}}{1+\sqrt{1+4\tilde\Lambda^2/m^2}}\right)+2\sqrt{1+4\tilde\Lambda^2/m^2}\right] = 0 .
\end{equation}
The BPS masses are satisfying $M_{(1,0)} = M_{(1,-1)} +
M_{(0,1)}$ on this curve, and thus $(n_e,n_D)=(1,0)$ as the bound state of
$(1,-1)$ and $(0,1)$ becomes unstable and decays into fundamental states
when the mass parameter is inside of the CMS.

We have discussed the correspondence between quantum Hall state as the incompressible fluid and vortex state.
Thus elementary excitations on a non-Abelian vortex is investigated in this section \ref{world_sheet}.
Then we will consider a physical interpretation of these elementary excitations in terms of quantum Hall ferromagnets in the following section.

\section{Particle states in quantum Hall ferromagnets}\label{sec:CP}

At last, we now discuss a relationship between the supersymmetric $\mathbb{C}{\bf P}^1$ model and the quantum Hall ferromagnets.
Let us start with the meaning of the kink solution of the $\mathbb{C}{\bf P}^1$ model in the context of the quantum Hall state.
According to the internal degree of freedom, two isolated vacua appear in the vortex theory.
Since this internal space $\mathbb{C}{\bf P}^1 = SU(2)/U(1)$ corresponds to the spin degree of the electron, the $\mathbb{C}{\bf P}^1$ coordinate is interpreted as the spin or pseudo spin direction.
Thus the kink excitation interpolates two polarized states.
In terms of the bilayer quantum Hall system, these two vacua correspond to the top and bottom layers.
Therefore a kinked vortex $(n_e,n_D)=(0,1)$ is a magnetic flux penetrating two layers and $(-1,1)$ excitation has also an electric charge.
This is an electron attached with a flux which is namely a composite particle state\cite{PhysRevLett.48.1144,PhysRevLett.49.957}.
We propose that a pure
 electron state is absence from the strong coupling region but only a
 composite particle state is present.

\begin{figure}[htbp]
 \begin{center}
  \includegraphics[height=2.3cm]{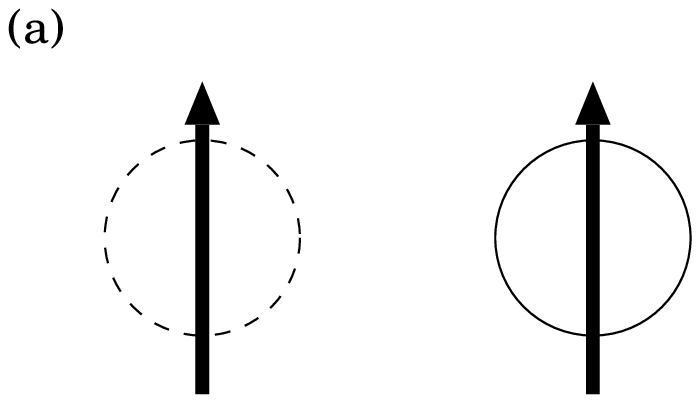}
  \hspace{1cm}
  \includegraphics[height=2.3cm]{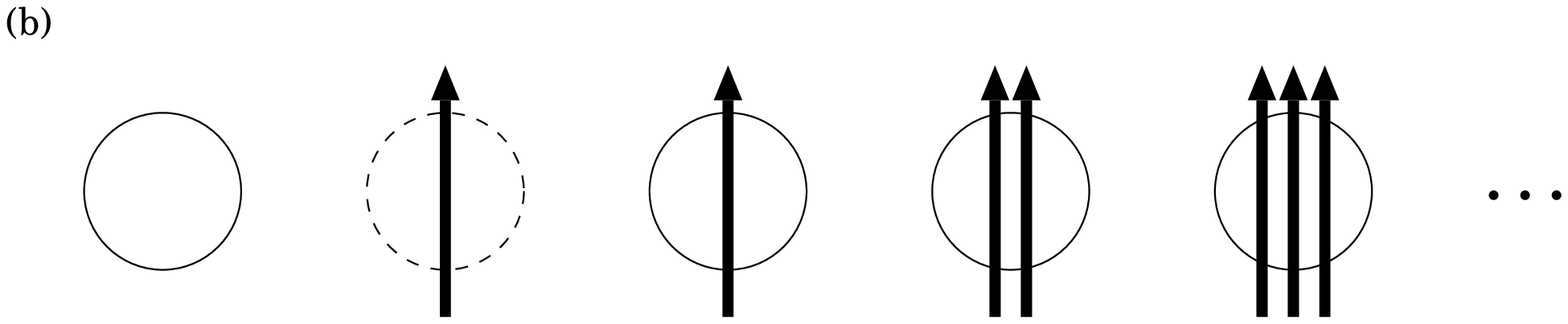}
 \end{center}
 \caption{Particle states as elementary excitations in the vortex theory
 (a) Charged particle (electron) must accompany a magnetic flux in the
 small $m^2$ region. (b) Pure electron state can be observed in the
 large $m^2$ region.} 
 \label{composite}
\end{figure}

Then the kink excitation is obtained by the dimensional reduction of the monopole in the four dimensional theory\cite{Hanany:2004ea}.
The singularity of the monopole, which is the Dirac string, can transmute the statistics of the particle.
As a result of the $\mathbb{C}{\bf P}^1$ model, permitted vortex states
are only the penetrating magnetic flux and the composite particle state,
and the pure electron state is forbidden in the small $m^2$
region[Fig.\ref{composite}(a)]. 

In the large $m^2$ region, however, the restriction of elementary excitations is soften.
That means the stability of the composite particle becomes ambiguous 
and a pure electric excitation $n_e=1$, $n_D=0$ can be observed[Fig.\ref{composite}(b)].
This excitation is considered as the bound state of the kink $n_e=0$, $n_D=1$ and the opposite composite particle $n_e=1$, $n_D=-1$ generated by fluctuations.
Thus this fact suggests the small $m^2$ region is the quantum Hall phase
but the quantum Hall ferromagnet state as the non-Abelian vortex fluid state is broken as the twisted mass becomes larger.
Furthermore, by the analysis of the $\mathbb{C}{\bf P}^1$ model, the CMS giving the phase transition point is exactly evaluated.
However, we have applied the $1+1$ dimensional model to the vortex theory
although the vortex is not elongated infinitely.
Thus the kink solution spectrum can be corrected by finite size effect.

To confirm the above discussion, we should discuss a meaning of the twisted mass parameter in terms of the quantum Hall ferromagnets.
In the classical limit where the effect of the dynamical mass can be neglected, the two vacua are obtained as $\sigma_\pm \simeq \pm m/2$, and the spectrum of the kink excitation becomes $M \simeq m$.
Because the kink spectrum is considered as difference of energy levels between the top and bottom layers, we now identify the twisted mass parameter $m$ with bias gate voltage between layers $V_{\mathrm{bias}}$, which corresponds to Zeeman energy in the context of the spin system.

Furthermore the twisted mass includes the imaginary part induced by the
theta angle $\vartheta$. 
When the mass is pure imaginary at $\vartheta=\pi$, the topological mass
$m_D$ vanishes, and thus the vortex worldsheet obtains superconformal symmetry. 
In general, this superconformal point of the $\mathbb{C}{\bf P}^{N-1}$
model at $m_k =- \exp (2\pi i k/N) \tilde\Lambda$ for $k=1, \cdots, N$
is related to the $A_{N-1}$ series of the $ADE$
classification\cite{Witten:1993jg}. 
At the critical point, the real part of the twisted mass vanishes, and
this fact suggests that the bias gate also vanishes. 
Since $\vartheta$ is originally considered as the mixing angle of the
electric and magnetic field,  
the electric field is fully converted to magnetic at $\vartheta= \pi$.
Another considerable interpretation is that $\vartheta/2$ is a tilt angle of
the external magnetic field from the two dimensional layers.
This means the external field is perpendicular to the layers
at $\vartheta = 0$, and is parallel at $\vartheta = \pi$.
However this interpretation is a little confusing because the
incompressible fluid can be hardly constructed with the parallel external field.
On the other hand, in the case of the $n$-vector model, the twisted mass
is associated with the coupling constant, $J/kT = \tilde \Lambda /
m$\cite{hikami2007ftc}.  
According to the imaginary part of the coupling, the corresponding Hamiltonian becomes
non-hermitian operator. 
In this sense, we might apply superconformal field theory to non-hermitian quantum mechanics.

Finally we mention the filling fraction of the quantum Hall state.
When the numbers of the kink excitations $(n_e, n_D)=(0,1)$ and the
composite states $(-1,1)$ are $N_f$ and $N_c$ respectively, the filling
fraction which is the ratio of the particles and the total fluxes is
represented as 
\begin{equation}
 \nu = \frac{N_c}{N_f + N_c}.
\end{equation}
A disappointing part of our analysis is that we mainly focus on each
vortex independently because we treat the interactionless sector of the
vortex fluid state.
It is 1-body problem. 
Thus we cannot discuss the ratio of the elementary excitations more.
On the other hand, according to the correspondence between the
noncommutative parameter $\theta=1/(2\pi \rho_e)=1/(B\nu)$ and the FI
parameter or the coupling constant of the $\mathbb{C}{\bf P}^1$ model
$r=2/g^2$, we can associate the filling fraction with the twisted mass
parameter from the relation (\ref{dynamical_mass}), $\nu \approx \pi/(B\log(m/\Lambda))$.
Although this estimation is at the classical level and available in the
weak coupling region,  
this means the filling fraction decreases as the mass becomes larger,
and supports the breakdown of our description for the quantum Hall state
as the previous discussion. 
In the large $m^2$ region where the bias gate voltage becomes larger,
it seems the bilayer system is decoupled, and hence the non-Abelian vortex fluid
is not good description for the quantum Hall ferromagnets.

\section{Discussions}\label{sec:discussion}

In this paper, we have discussed the incompressible fluid as the LLL
state and its effective theory.
The background magnetic field induces the spatial noncommutativity, but
it is realized by the infinite dimensional Hilbert space where the
number of particles is infinite.
This infiniteness is regularized by the boundary field modifying the
commutation relation.
As a result, one obtains the matrix Chern-Simons theory as the effective
theory of the regularized incompressible
fluid\cite{polychronakos2001qhs,polychronakos2001uqhs}. 
Thus the phase space of the incompressible fluid is spanned by finite
dimensional matrix variables satisfying the modified commutation
relation.  

On the other hand, the modified commutation relation is also observed in
the vortex theory.
It is the supersymmetric vacuum condition for the vortex theory, and
characterizing the moduli space of vortices\cite{hanany2003via}.
Thus the incompressible fluid is considered as the vortex fluid state.
This correspondence suggests that the phase space of the incompressible
fluid is equivalent to the moduli space of the vortex.

The non-Abelian vortex possesses the internal space
$\mathbb{C}{\bf P}^{N-1}$.
Indeed the symmetry of the quantum Hall ferromagnet $SU(N)$ is
decomposed to the electric charge part $U(1)$ and the spin part
$\mathbb{C}{\bf P}^{N-1}$. 
This decomposition is called the spin-charge separation.
Thus the non-Abelian vortex fluid state is considered as the
quantum Hall ferromagnet.
According to this correspondence, particle state of the quantum Hall
ferromagnet is investigated by the vortex theory.

To study the vortex state, we have applied the supersymmetric
$\mathbb{C}{\bf P}^{N-1}$ model to the vortex world-sheet
theory\cite{Hanany:2004ea}. 
In the case of the $\mathbb{C}{\bf P}^1$ model, the mass spectrum of
the topological excitation, the mass of the kink solution and the CMS
which is the marginal line of the strong and weak coupling region on the
complex twisted mass space are exactly evaluated. 
This twisted mass $m$ characterizes the mass scale of the kink
excitation, and thus it is considered as the bias gate voltage between
the top and bottom layers in the case of the bilayer quantum Hall system.

The elementary excitations in the strong coupling, equivalently the
small $m^2$ region, are only $(n_e, n_D)=(0,1)$ and $(1,-1)$ modes, and this
result implies only the composite particle appears, and the pure
electron state is forbidden in this region.
On the other hand, the composite particle is decomposed to the pure
electron and the magnetic flux in the large $m^2$ region.
Thus it proposes the phase transition between the strong
coupling region which is the non-Abelian vortex fluid phase and the weak
coupling region where the non-Abelian vortex description is not available.
Therefore the CMS separating the two phases gives the transition line for the
breakdown of the non-Abelian vortex description of the quantum Hall
ferromagnet.

Then we now comment some issues of our approach in perspective.
In the quantum Hall state, the edge excitation plays an important role on the transport phenomena, and the edge state is well described by conformal field theory, induced on boundary of a manifold on which Chern-Simons theory is defined.
This is an example of the holographic relation of the bulk/edge duality.
In the case of Chern-Simons matrix
theory\cite{polychronakos2001qhs,polychronakos2001uqhs}, one obtain the
one dimensional quantum many-body model, which is called Calogero
model\cite{calogero:2197}, or Sutherland model\cite{Sutherland:1971ic}.
Thus it is expected that Calogero model with internal degrees of freedom is obtained from the quantum Hall ferromagnets.
Furthermore, it is well known that the quantized filling fraction possesses the hierarchical structure, which has been discussed in the context of the matrix model\cite{Cappelli:2006wa}.
The elementary excitation of the vortex in the hierarchical state should be understood.










\subsection*{Acknowledgments}

The author would like to thank S. Hikami for stimulating conversations
and reading the manuscript. 
The author also thank H. Shimada and T. Yoshimoto for useful
discussions, G. Marmorini and M. Nitta for valuable comments.


\begin{thebibliography}{10}

\bibitem{prange1990qhe}
R.~Prange and S.~Girvin, {\em {The quantum Hall effect.}}
\newblock Springer Verlag New York, 1990.

\bibitem{PhysRevB.44.5246}
A.~Lopez and E.~Fradkin, {\it {Fractional quantum Hall effect and Chern-Simons
  gauge theories}},  {\em Phys. Rev. B} {\bf 44} (1991) 5246.

\bibitem{seiberg1999sta}
N.~Seiberg and E.~Witten, {\it {String theory and noncommutative geometry}},
  {\em JHEP} {\bf 9909} (1999) 026,
  [\href{http://arxiv.org/abs/hep-th/9908142}{{\tt hep-th/9908142}}].

\bibitem{susskind2001qhf}
L.~Susskind, {\it {The quantum Hall fluid and non-commutative Chern-Simons
  theory}},  \href{http://arxiv.org/abs/hep-th/0101029}{{\tt hep-th/0101029}}.

\bibitem{polychronakos2001qhs}
A.~Polychronakos, {\it {Quantum Hall states as matrix Chern-Simons theory}},
  {\em JHEP} {\bf 0104} (2001) 009,
  [\href{http://arxiv.org/abs/hep-th/0103013}{{\tt hep-th/0103013}}].

\bibitem{polychronakos2001uqhs}
A.~Polychronakos, {\it {Quantum Hall states on the cylinder as unitary matrix
  Chern-Simons theory}},  {\em JHEP} {\bf 0106} (2001) 056,
  [\href{http://arxiv.org/abs/hep-th/0106011}{{\tt hep-th/0106011}}].

\bibitem{hanany2003via}
A.~Hanany and D.~Tong, {\it {Vortices, instantons and branes}},  {\em JHEP}
  {\bf 0307} (2003) 031, [\href{http://arxiv.org/abs/hep-th/0306150}{{\tt
  hep-th/0306150}}].

\bibitem{tong2004qhf}
D.~Tong, {\it {A quantum Hall fluid of vortices}},  {\em JHEP} {\bf 0402}
  (2004) 038, [\href{http://arxiv.org/abs/hep-th/0306266}{{\tt
  hep-th/0306266}}].

\bibitem{dorey1998bst}
N.~Dorey, {\it {The BPS spectra of two-dimensional supersymmetric gauge
  theories with twisted mass terms}},  {\em JHEP} {\bf 9811} (1998) 005,
  [\href{http://arxiv.org/abs/hep-th/9806056}{{\tt hep-th/9806056}}].

\bibitem{shifman2006cca}
M.~Shifman, A.~Vainshtein, and R.~Zwicky, {\it {Central charge anomalies in 2D
  sigma models with twisted mass}},  {\em J. Phys.} {\bf A39} (2006) 13005,
  [\href{http://arxiv.org/abs/hep-th/0602004}{{\tt hep-th/0602004}}].

\bibitem{Hanany:2004ea}
A.~Hanany and D.~Tong, {\it {Vortex strings and four-dimensional gauge
  dynamics}},  {\em JHEP} {\bf 0404} (2004) 066,
  [\href{http://arxiv.org/abs/hep-th/0403158}{{\tt hep-th/0403158}}].

\bibitem{Hanany:1997vm}
A.~Hanany and K.~Hori, {\it {Branes and ${\cal N} = 2$ theories in two
  dimensions}},  {\em Nucl. Phys.} {\bf B513} (1998) 119,
  [\href{http://arxiv.org/abs/hep-th/9707192}{{\tt hep-th/9707192}}].

\bibitem{hikami2007ftc}
S.~Hikami and T.~Yoshimoto, {\it {Instanton and superconductivity in
  supersymmetric CP$(N-1)$ model}},  {\em J. Phys.} {\bf A40} (2007) F369,
  [\href{http://arxiv.org/abs/cond-mat/0703691}{{\tt cond-mat/0703691}}].

\bibitem{atiyah1978cip}
M.~Atiyah, N.~Hitchin, V.~Drinfeld, and Y.~Manin, {\it {Construction of
  instantons}},  {\em Phys. Lett.} {\bf A65} (1978) 185.

\bibitem{Taubes:1979tm}
C.~H. Taubes, {\it {Arbitrary N: Vortex Solutions to the First Order Landau-
  Ginzburg Equations}},  {\em Commun. Math. Phys.} {\bf 72} (1980) 277.

\bibitem{Eto:2005yh}
M.~Eto, Y.~Isozumi, M.~Nitta, K.~Ohashi, and N.~Sakai, {\it {Moduli space of
  non-Abelian vortices}},  {\em Phys. Rev. Lett.} {\bf 96} (2006) 161601,
  [\href{http://arxiv.org/abs/hep-th/0511088}{{\tt hep-th/0511088}}].

\bibitem{Auzzi:2005gr}
R.~Auzzi, M.~Shifman, and A.~Yung, {\it {Composite non-Abelian flux tubes in $N
  = 2$ SQCD}},  {\em Phys. Rev.} {\bf D73} (2006) 105012,
  [\href{http://arxiv.org/abs/hep-th/0511150}{{\tt hep-th/0511150}}].

\bibitem{eto:065021}
M.~Eto, K.~Konishi, G.~Marmorini, M.~Nitta, K.~Ohashi, W.~Vinci, and N.~Yokoi,
  {\it {Non-Abelian vortices of higher winding numbers}},  {\em Phys. Rev.}
  {\bf D74} (2006) 065021, [\href{http://arxiv.org/abs/hep-th/0607070}{{\tt
  hep-th/0607070}}].

\bibitem{PhysRevLett.48.1144}
F.~Wilczek, {\it {Magnetic Flux, Angular Momentum, and Statistics}},  {\em
  Phys. Rev. Lett.} {\bf 48} (1982) 1144.

\bibitem{PhysRevLett.49.957}
F.~Wilczek, {\it {Quantum Mechanics of Fractional-Spin Particles}},  {\em Phys.
  Rev. Lett.} {\bf 49} (1982) 957.

\bibitem{Witten:1993jg}
E.~Witten, {\it {On the Landau-Ginzburg description of N=2 minimal models}},
  {\em Int. J. Mod. Phys.} {\bf A9} (1994) 4783--4800,
  [\href{http://arxiv.org/abs/hep-th/9304026}{{\tt hep-th/9304026}}].

\bibitem{calogero:2197}
F.~Calogero, {\it {Ground State of a One-Dimensional $N$-Body System}},  {\em
  J. Math. Phys.} {\bf 10} (1969) 2197.

\bibitem{Sutherland:1971ic}
B.~Sutherland, {\it {Quantum many body problem in one-dimension: Ground
  state}},  {\em J. Math. Phys.} {\bf 12} (1971) 246.

\bibitem{Cappelli:2006wa}
A.~Cappelli and I.~Rodriguez, {\it {Jain states in a matrix theory of the
  quantum Hall effect}},  {\em JHEP} {\bf 0612} (2006) 056,
  [\href{http://arxiv.org/abs/hep-th/0610269}{{\tt hep-th/0610269}}].

\end{thebibliography}

\providecommand{\href}[2]{#2}\begingroup\raggedright\endgroup

\end{document}